%% file: main.tex
\documentclass[acmsmall,screen]{acmart}

\usepackage{hyperref}
\usepackage{subcaption}
\usepackage{siunitx}
\usepackage{tabularx}
\usepackage[table]{xcolor}
\usepackage[]{collab}
\usepackage{algorithm}
\usepackage{algpseudocode}
\usepackage{multirow}
\usepackage{tikz}
\usepackage{amsmath}
\usepackage{xspace}
\usepackage{enumitem}
\usepackage{booktabs}
\usepackage{bbding}
\usepackage{stfloats}
\usepackage{indentfirst}
\usepackage{makecell}
\usepackage[font=small]{caption}
\usepackage{tcolorbox}
\usepackage{lipsum}
\usepackage{minted}
\usepackage{graphicx}
\usepackage[normalem]{ulem}
\usepackage{minted} 
\usepackage{multicol}
\usepackage{enumitem}
\usepackage{graphicx}
\usepackage{subcaption}
\usepackage{array}
\usepackage{ragged2e}
\usepackage{tabularx} 
\usepackage{color}
\usepackage{colortbl}
\definecolor{graybg}{gray}{0.95}

\collabAuthor{zb}{teal}{Zhuangbin Chen}

\newcommand{\ie}{{\em i.e.},\xspace}
\newcommand{\eg}{{\em e.g.},\xspace}

\newcommand{\combench}{\textsc{Com}\textsc{Bench}\xspace}

\AtBeginDocument{%
  }

\setcopyright{acmlicensed}
\copyrightyear{2026}
\acmYear{2026}
\acmDOI{XXXXXXX.XXXXXXX}
\acmConference[ISSTA '26]{ACM SIGSOFT International Symposium on Software Testing and Analysis}{October 3--9, 2026}{Oakland, California, USA}
\acmISBN{978-1-4503-XXXX-X/2018/06}

\begin{document}

\title[A Repository-level Real-world Benchmark for Compilation Error Repair]{ComBench: A Repository-level Real-world Benchmark for Compilation Error Repair}

\author{Jia Li}
\email{linsayli@link.cuhk.edu.hk}
\orcid{0009-0007-5557-4916}
\affiliation{%
  \institution{The Chinese University of Hong Kong}
  \city{Hong Kong}
  \country{China}
}

\author{Zeyang Zhuang}
\email{zyzhuang22@cse.cuhk.edu.hk}
\affiliation{%
  \institution{The Chinese University of Hong Kong}
  \city{Hong Kong}
  \country{China}
}

\author{Zhuangbin Chen}
\email{chenzhb36@mail.sysu.edu.cn}
\affiliation{%
  \institution{School of Software Engineering, Sun Yat-sen University}
  \city{Zhuhai}
  \country{China}
}

\author{Yuxin Su}
\email{suyx35@mail.sysu.edu.cn}
\affiliation{%
  \institution{Sun Yat-sen University}
  \city{Guangzhou}
  \country{China}
}

\author{Wei Meng}
\email{wei@cse.cuhk.edu.hk}
\affiliation{%
  \institution{The Chinese University of Hong Kong}
  \city{Hong Kong}
  \country{China}
}

\author{Michael R. Lyu}
\email{lyu@cse.cuhk.edu.hk}
\affiliation{%
  \institution{The Chinese University of Hong Kong}
  \city{Hong Kong}
  \country{China}
}




\input{sections/00-abstract}
\maketitle
\input{sections/01-introduction}
\input{sections/02-background}

\input{sections/03-method}

\input{sections/04-analysis}
\input{sections/05-evaluation}
\input{sections/06-conclusion}

\bibliographystyle{ACM-Reference-Format}
\bibliography{main}

\appendix

\end{document}

%% file: sections/00-abstract.tex
\begin{abstract}
Compilation errors pose pervasive and critical challenges in software development, significantly hindering productivity.
Therefore, Automated Compilation Error Repair (ACER) techniques are proposed to mitigate these issues. Despite recent advancements in ACER, its real-world performance remains poorly evaluated. This can be largely attributed to the limitations of existing benchmarks, \ie decontextualized single-file data, lack of authentic source diversity, and biased local task modeling that ignores crucial repository-level complexities. To bridge this critical gap, we propose \combench, the first repository-level, reproducible real-world benchmark for C/C++ compilation error repair.

\combench is constructed through a novel, automated framework that systematically mines real-world failures from the GitHub CI histories of large-scale open-source projects. Our framework contributes techniques for the high-precision identification of ground-truth repair patches from complex version histories and a high-fidelity mechanism for reproducing the original, ephemeral build environments. To ensure data quality, all samples in \combench are execution-verified---guaranteeing reproducible failures and build success with ground-truth patches. Using \combench, we conduct a comprehensive evaluation of 12 modern LLMs under both direct and agent-based repair settings. Our experiments reveal a significant gap between a model's ability to achieve syntactic correctness (a 73\% success rate for GPT-5) and its ability to ensure semantic correctness (only 41\% of its patches are valid). We also find that different models exhibit distinct specializations for different error types. \combench provides a robust and realistic platform to guide the future development of ACER techniques capable of addressing the complexities of modern software development.
\end{abstract}

%% file: sections/01-introduction.tex
\section{Introduction}\label{sec:introduction}

Compilation errors~\cite{compileerror} are a pervasive and disruptive challenge within the core ``edit-compile-debug'' software development cycle.
According to a large-scale study~\cite{10.1145/2568225.2568255}, C++ developers experience compilation failures in 37.4\% of their 10.1 daily builds, while Java developers face a 29.7\% failure rate across 6.98 daily builds. Resolving these errors manually is a time-consuming and tedious task, with reported median repair times of 5 mins for C++ and 12 mins for Java, exhibiting an order of magnitude variation across error types. This forces programmers to switch from feature development to troubleshooting, leading to blockages and delays that impact overall productivity.

\textit{Automated compilation error repair} (ACER) techniques~\cite{10.5555/3298239.3298436,10.1145/3338906.3340455,10.1145/3551349.3560422,10589843,rust} are proposed to alleviate these issues.
The advent of LLMs has significantly enhanced the effectiveness of code generation, which boosts the development of LLM-based ACER tools~\cite{rust}.
However, fully assessing their performance is challenging due to a lack of benchmarks reflecting real-world compilation error complexity.
Existing ACER benchmarks~\cite{10.5555/3298239.3298436,8445185} deviate significantly from the practices of real-world software development across three dimensions: \textit{data granularity}, \textit{source authenticity}, and \textit{task modeling}.
First, their decontextualized, single-file paradigm (typically around 20 lines) isolates errors from their project environment, preventing the capture of essential repository-level complexities, such as cross-file references and inter-module dependencies.
Second, their narrow data sources have led to a lack of authenticity and diversity.
Existing benchmarks are exclusively derived from introductory students' programming assignments. This results in a relatively homogeneous set of error types (\eg missing delimiters).
Thus, they fail to represent the complex and diverse errors in large-scale projects.
Lastly, this oversimplified data setting leads to biased task models that frame ACER as a local, sequence-to-sequence problem.
It overlooks critical challenges such as global fault localization and comprehensive context collection within large repositories.
Consequently, existing benchmarks create a significant disconnect between academic research and industrial practice. This highlights the urgent need for a more comprehensive and representative benchmark.

To address these limitations, we propose \combench, a repository-level real-world benchmark for compilation error repair with reproducible environments and developer-provided ground-truth fixes. \combench aims to provide a unified platform for more accurately evaluating the performance of ACER methods. It features several distinct advantages. (1) \textit{Repository-level Context:} \combench's data is directly sourced from Git repository histories of open-source projects. Each error instance includes the complete project-level context, \ie all project code, file structure, and dependencies. This enables repair methods to consider cross-file context.
(2) \textit{Authentic Real-world Errors}: Compilation errors are collected from GitHub's Continuous Integration (CI) system~\cite{ciwiki}, GitHub Actions~\cite{actions}. Error-inducing code and its corresponding repair patches are derived from actual developer commits, ensuring the authenticity and representativeness of the data.
(3) \textit{Comprehensive Error Types with Labels}: The benchmark encompasses a wide variety of compilation error types, from simple syntax to complex cross-file issues. Each error instance includes detailed compiler diagnostics, complete code context, and the original developer's repair patch, which is essential for evaluating repair intent and correctness.
(4) \textit{Reproducible and Continuously Evolving}: \combench provides Docker containers for consistent compilation environments, ensuring reproducibility of reported errors. Furthermore, a reusable data collection pipeline ensures continuous updates, allowing the benchmark to evolve with new errors and development practices from GitHub.

The construction of such a benchmark presents three significant challenges.
First, \textit{the heterogeneity and noise in CI logs} make it difficult to precisely extract structured compiler diagnostics. These logs lack a uniform format and are intermingled with massive irrelevant information, such as environment preparation, package installation, test results, and deployment details.
Second, a more complex challenge lies in \textit{the causal inference of the ground truth repair}, specifically locating the exact fix commit and isolating its minimal repair patch. This task is intricate because compilation failures often occur on ephemeral feature branches, complicating the direct traceability of fix commits and necessitating sophisticated search algorithms. Furthermore, identified fix commits frequently bundle the actual error repair with extensive, unrelated modifications like new features or refactoring. Precisely extracting the minimal, targeted patch from these multi-purpose changes is crucial for benchmark quality but inherently difficult.
Finally, to ensure the benchmark’s usability for evaluation, the critical challenge is \textit{reproduction}: accurately recreating the exact historical CI environment in which the failure originally occurred. This is often complicated by two factors, \ie inaccessible proprietary runner images, which prevent proper container runtime environment setup, and ambiguous job configurations lacking essential execution parameters, thus hindering a precise reconstruction of the job's specific runtime environment.

Our \combench framework systematically addresses these challenges through a targeted, multi-faceted approach. To identify compilation errors from noisy and heterogeneous CI logs, we employ a dual-filtering algorithm to prune irrelevant entries and validate error lines, complemented by a bidirectional context collector for detailed error information extraction.
For causally inferring the ground truth repair, \combench utilizes a multi-source search algorithm to locate and rank candidate repair commits across all branches. Following this, compilation dependency analysis precisely isolates the minimal repair patch by filtering for only the modified files directly implicated in resolving the error, thereby ensuring high-quality ground truth.
Finally, we ensure the benchmark's reproducibility for evaluation through two mechanisms. For inaccessible proprietary runner images, a two-layer rebuild process maps the original build environment to a generic Docker image and then overlays project-specific pre-building commands. Concurrently, an intelligent matching algorithm addresses ambiguous job configurations by inferring job-specific parameters from human-readable names, enabling accurate reconstruction of the historical CI environment.
To ensure the correctness of collected data, we perform end-to-end reproducible validation for each candidate instance: we reproduce the compilation failure in the reconstructed environment and verify that applying the developer fix makes the build succeed (i.e., fail-fix-pass). We retain only instances that pass this check; all 200 released instances satisfy it.


Our constructed benchmark, \combench, comprises 200 C/C++ compilation errors, encompassing seven major categories of compilation issues. These include prevalent declaration, type, member, and syntax errors, as well as less frequent function, template, and semantic errors. These errors are sourced from four prominent large-scale C/C++ open-source projects, \ie OpenSSL~\cite{openssl}, Bitcoin~\cite{bitcoin}, RocksDB~\cite{rocksdb}, and LLVM~\cite{llvm}. They span diverse application domains, \ie security, blockchain, database management, and compiler infrastructure, thereby ensuring \combench's broad representativeness of real-world development scenarios and the diverse error types.

To rigorously evaluate the effectiveness of current LLM-based ACER techniques against these real-world compilation errors, we conduct comprehensive experiments involving 12 popular LLMs of varying parameter scales. We explore two operational settings: \textit{direct repair with perfect localization}, which assesses an LLM's core patch generation capabilities by abstracting away the complex task of fault localization; and \textit{agent-based repair}, which evaluates the LLM's ability to autonomously reason, localize, and iteratively refine repairs.
Beyond mere \textit{compilation success}, our evaluation metrics are designed to provide a nuanced understanding of repair quality. We include \textit{semantic correctness} to ensure that any generated fix preserves the original program's intended functionality, and \textit{exact match} with ground-truth developer fixes to assess alignment with human-expert solutions.
Our experiments reveal the varying capabilities of these LLMs for the ACER task, highlight crucial limitations in current agent designs when faced with real-world complexities, and provide an in-depth analysis of their repair performance across different error types. 
Specifically, GPT-5~\cite{openai_gpt5} consistently demonstrates the highest overall repair capability, while Gemini-2.5-Pro~\cite{google_gemini} exhibits superior semantic understanding capability. Agent-based approaches showed higher semantic correctness compared to direct repair; however, current designs still require further improvements.

In summary, this work makes the following major contributions:

\begin{itemize}[noitemsep,leftmargin=5.5mm]
    \item We introduce \combench, the first repository-level, reproducible real-world benchmark for compilation error repair. \combench addresses critical limitations of existing datasets by offering authentic errors sourced from GitHub CI with full project context, comprehensive error types, and rich annotations including ground-truth developer fixes.
    \item We develop a novel, robust framework for benchmark construction, which systematically overcomes challenges in collecting real-world compilation errors. This includes techniques for precise error extraction from noisy CI logs, causal inference of minimal repair patches from complex Git histories, and accurate reproduction of historical build environments.
    \item We conduct a comprehensive empirical evaluation of 12 popular LLMs based on \combench. Our evaluation reveals varying LLM capabilities, highlights limitations in current agent designs, and provides novel insights into their repair performance across diverse error types.
\end{itemize}

%% file: sections/02-background.tex
\section{Background}

This section discusses existing research to contextualize our work. We first introduce the task of ACER and its challenges.
Next, we review mainstream ACER methods and present the inherent limitations of existing benchmarks, thereby justifying the necessity of constructing \combench. Finally, we introduce GitHub Actions~\cite{actions}, an open-source CI platform where repository-level, real-world compilation errors can be collected.

\subsection{Automated Compilation Error Repair} \label{sec:benchmarks}

Automated Program Repair (APR)~\cite{10.1145/3696450} aims to automatically fix software defects. It generally targets three types of issues: logical errors (bugs)~\cite{10.1145/3715754,ruan2025specrover,zhang2024autocoderover,yin2024thinkrepair, xia2024automated}, security vulnerabilities~\cite{fu2022vulrepair,10.1145/3597503.3639222,zhou2025large,zhang2024vuladvisor}, and compilation errors~\cite{10.5555/3298239.3298436,10.1145/3338906.3340455,10.1145/3551349.3560422,rust}.
Among these, logical errors occur when a program's behavior deviates from its functional specification; these errors are typically detected through test failures. Security vulnerabilities are a specific type of logical error that can be exploited for malicious purposes, and they are often identified by security analysis tools. In contrast, compilation errors, which are the specific focus of ACER, arise when a program violates the syntax or static type rules of a programming language, thereby preventing its successful build.
Unlike logical or security issues, compilation errors are detected instantly and deterministically by the compiler. The compiler provides rich, structured diagnostic information, including error codes, locations, and descriptive messages. 
Since successful compilation is a prerequisite for all subsequent development, testing, and deployment stages, the automated and efficient repair of these errors is crucial for maintaining a smooth development workflow and boosting overall productivity. 
Moreover, their deterministic nature and frequent occurrence make them an ideal and high-return target for automated repair.

The core task of ACER is to automatically generate a fix patch for a non-compiling program, utilizing the available compiler diagnostics (\eg error location and messages), such that the program not only compiles successfully but also retains its intended semantics. However, despite the valuable diagnostic information provided by compilers, automatically generating a semantically correct and robust fix patch presents several significant challenges:

\begin{itemize}[leftmargin=5.5mm, labelsep=0.5em, itemsep=0.5ex, topsep=0.5ex, parsep=0ex]
    \item \textit{Symptom vs. Root Cause}: Compiler diagnostics often pinpoint symptoms rather than the underlying root cause. The actual problem might originate in a different part of the codebase than the reported error, necessitating a global understanding for accurate fault localization.

    \item \textit{Ambiguity in Repair Strategies}: A single diagnostic message can correspond to multiple syntactically plausible repair strategies. Discerning between these and selecting the one that preserves intended semantics requires a deeper, global contextual understanding of the code's structure, data flow, and control flow.

    \item \textit{Diverse Error Taxonomy}: Compilation errors encompass a wide spectrum of issues (\eg syntax errors, type mismatches, missing declarations). Each category often demands a distinct and tailored repair strategy, challenging the wide applicability of ACER tools.

    \item \textit{Global Consistency}: A seemingly correct local fix might introduce new compilation errors or semantic deviations elsewhere. The challenge of maintaining global consistency and preventing cascading failures makes generating a robust and globally sound patch highly complex.
\end{itemize}

\subsection{Compilation Error Repair Methods}

In recent years, deep learning-based methods have become the main approach for ACER. These techniques leverage the powerful pattern recognition and sequence modeling capabilities of neural networks to automatically generate fix patches by learning from vast datasets of error-fix pairs. Their evolution reveals a clear path of continuously enriching input information, optimizing model architectures, and innovating problem modeling. Early approaches, such as DeepFix~\cite{10.5555/3298239.3298436}, established a sequence-to-sequence foundation. Subsequent methods, including DeepDelta~\cite{10.1145/3338906.3340455} and TransRepair~\cite{10.1145/3551349.3560422}, integrated compiler feedback and code context. More recently, LabelRepair~\cite{10589843} optimized the process through sequence labeling. Recent work, such as RUSTASSISTANT~\cite{rust}, has introduced an iterative repair loop that repeatedly queries an LLM and uses the resulting compiler feedback to progressively refine patches. All these learning-based methods evolve to progressive improvements in repair accuracy.

However, the architectures and capabilities of these learning-based methods are heavily influenced and constrained by existing datasets and benchmarks~\cite{10.5555/3298239.3298436, 8445185}. Since these methods are primarily trained and evaluated on single-file and context-limited datasets, they inherently only prioritize local fixes. As a result, their ability to handle cross-file dependencies and project-level compilation issues remains largely unexplored. This highlights how the evaluation dataset significantly shapes method design and underscores the necessity for new, more representative benchmarks.

Current mainstream benchmarks for ACER exhibit notable limitations in capturing the complexity of real-world development scenarios, which is revealed by significant discrepancies from real-world software development practices across three key aspects:

\begin{itemize}[leftmargin=5.5mm, labelsep=0.5em, itemsep=0.5ex, topsep=0.5ex, parsep=0ex]
    \item \textit{Data Granularity and Context}: Existing benchmarks, such as DeepFix and TRACER, adopt a decontextualized, single-file paradigm, providing only short code snippets (\eg avg. under 20 lines). This approach isolates error instances from their project context, thereby failing to capture repository-level complexities essential for real-world repair, such as cross-file symbol resolution, inter-module linking issues, and complex build system configurations.
    
    \item \textit{Data Source Realism and Diversity}: Current benchmarks suffer from limited data sources, which are exclusively constructed from novice student programming assignments (\eg DeepFix~\cite{10.5555/3298239.3298436}, TRACER~\cite{8445185}).
    This narrow origin leads to a homogeneous set of relatively simple error types (\eg missing delimiters or declarations).
    Thus, they fail to represent the diverse and complex errors prevalent in industrial software, which often stem from intricate system interactions in large-scale projects.
    Furthermore, DeepFix's evaluation benchmark notably lacks corresponding human-written ground truth patches, precluding the assessment of whether a proposed fix aligns with the developer’s intended corrections.

    \item \textit{Task Modeling Bias}: The aforementioned simplified data settings directly lead to a biased task modeling, where ACER is commonly framed as a local sequence-to-sequence problem~\cite{10.5555/3298239.3298436, 8445185,rust}. Models like DeepFix take the entire program as input~\cite{10.5555/3298239.3298436}. TRACER, in contrast, focuses on a single erroneous line~\cite{8445185}, and RUSTASSISTANT accepts isolated code snippets surrounding the locations in compiler diagnostics~\cite{rust}. Despite these varied input approaches, these models implicitly assume that all necessary information to solve the problem is locally available. This contrasts sharply with real-world scenarios, where the core challenge of repair involves global fault localization within a large repository and the integration of scattered contextual information. Even the authors of TRACER acknowledge that their model cannot effectively handle errors requiring global context~\cite{8445185}.
\end{itemize}

In summary, the interconnected limitations across data granularity, source authenticity, and task modeling create a significant gap between current ACER research and practical application. Consequently, repair models validated on existing benchmarks often have questionable generalization capabilities when applied to real, complex projects. Therefore, developing a repository-level benchmark, which addresses all these deficiencies, is crucial for advancing the field.

\subsection{GitHub Actions CI System}

GitHub Actions~\cite{actions} is a Continuous Integration/Continuous Delivery (CI/CD) platform that automates build, test, and deployment pipelines directly within repositories. It serves as a valuable source for collecting real-world project-level compilation errors. Its automation is defined using YAML files stored in the repository's \texttt{.github/workflows/} directory, which specify the events that trigger workflows, the jobs to be executed, and the individual steps within those jobs.

The system operates with a clear conceptual hierarchy. A \textit{Workflow} is the top-level unit, corresponding to an automated process, defined in a YAML file and initiated by repository events (\eg "push", "pull\_request"). Each workflow consists of one or more \textit{Jobs}. A job is a sequence of tasks that runs on a specific virtual machine (runner). Jobs can run in parallel or sequentially, and the build matrix (\texttt{strategy: matrix}) allows a single job definition to be expanded into multiple jobs under different configurations. Critically, the execution log generated by each job is a valuable source for collecting compilation errors. A \textit{Step} is the smallest unit of work within a job, executing sequentially. Steps can run shell commands or execute reusable \textit{Actions}, which encapsulate common tasks like checking out code (\texttt{actions/checkout}) or setting up toolchains (\texttt{actions/setup-python}).

However, this operational paradigm introduces a critical challenge: the \textit{opacity} of runner environments. Although logs capture the compilation output, the internal state of the virtual machine, including pre-installed system libraries, environment variables, and transient filesystem configurations, remains opaque to researchers. This environmental invisibility significantly complicates the reproduction of environment-dependent compilation failures, as the source code repository alone often fails to encapsulate the comprehensive context required for a successful build.

%% file: sections/03-method.tex
\section{\combench Construction}\label{sec:technique}

\noindent\textbf{Goal.}
Existing ACER benchmarks often under-approximate real-world compilation failures by (i) discarding repository context, (ii) lacking trustworthy ground truth, and (iii) being hard to reproduce. \combench is designed to close this gap by constructing \emph{execution-verified} \emph{repository-level} error--fix instances mined from real CI histories. Concretely, our construction must satisfy three requirements simultaneously: (1) \emph{authenticity} (errors and fixes originate from developers' commits), (2) \emph{ground truth} (a causally aligned fixing change is identified and minimized), and (3) \emph{reproducibility} (the original CI failure can be replayed locally under a faithful environment).

\begin{figure*}
    \centering
    \includegraphics[width=0.96\textwidth]{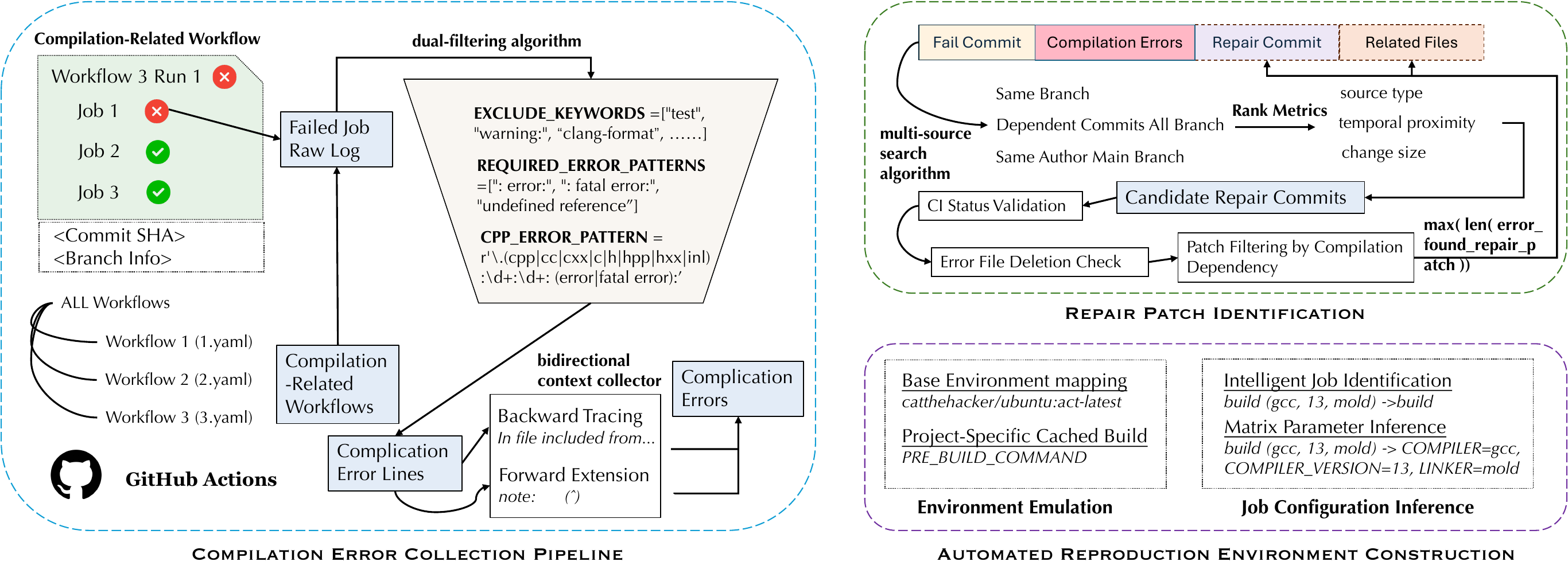}
    \caption{Overview of \combench construction workflow: (1) \textbf{Error Collection} via dual-filtering and context tracing; (2) \textbf{Patch Identification} through multi-source search and dependency-based filtering; and (3) \textbf{Environment Construction} using Docker-based emulation and job configuration inference for automated reproduction with \texttt{act}.}
    \label{fig:arch}
    \vspace{-10pt}
\end{figure*}

\noindent\textbf{Overview.}
As illustrated in Figure~\ref{fig:arch}, we build \combench in three stages. These stages can be grouped into two primary phases: (i) \textbf{data collection} (Stages 1--2), which follows a three-step pipeline of CI failure collection (Section~\ref{sec:ci_data_collect}), compilation error extraction (Section~\ref{sec:com_err_extra}), and repair patch identification (Section~\ref{sec:repair_patch}); and (ii) \textbf{reproduction environment construction} (Stage 3), which builds a high-fidelity local environment and performs end-to-end \textit{fail--fix--pass} execution verification (Section~\ref{sec:reproduce}). The resulting benchmark consists of reproducible, repository-level compilation errors, each paired with a developer-provided ground-truth fix. It facilitates the evaluation of global repair capability, which we assess using compilation success and intended-behavior consistency (Section~\ref{sec:exp-results}). We further analyze \combench to characterize its representativeness and coverage (Section~\ref{sec:statistic}).

\subsection{Benchmark Instance Definition and Design Constraints}
\label{sec:instance_def}

To make the construction objective explicit, we define a \combench instance as a tuple:
\[
\mathcal{I} = \langle \mathcal{R}, c_f, j_f, \mathcal{E}, c_r, \Delta_r, \mathcal{A} \rangle,
\]
where \(\mathcal{R}\) is a Git repository; \(c_f\) is the \emph{failure commit} that triggers a CI job \(j_f\) to fail; \(\mathcal{E}\) is the set of extracted compilation errors (each with compiler message, file/line, and contextual traces); \(c_r\) is the \emph{repair commit} that causally fixes the failure; \(\Delta_r\) is the \emph{minimal compilation-relevant patch} derived from \(c_r\); and \(\mathcal{A}\) is a reproduction artifact (Docker image, workflow, and inferred job parameters) that enables replaying the failure and validating fixes.

This definition highlights two key constraints that guide the rest of this section. First, \(c_r\) must be \emph{causally aligned} with \(c_f\), rather than being a later revision that happens to pass CI. Second, both the failure and the fix must be \emph{execution-verifiable} under \(\mathcal{A}\): applying \(\Delta_r\) to \(c_f\) should make the target job compile successfully.

\subsection{Data Source Selection}
The first step is selecting data sources that are likely to surface repository-level compilation failures. To ensure \combench is realistic and captures repository-level properties, we selected large-scale, actively-maintained C/C++ projects from GitHub~\cite{github}. We chose C/C++ for its prevalence and complex build processes, which yield diverse compilation failures, and GitHub for its complete version histories and CI build logs. We further refine candidate projects using three criteria: (1) substantial codebase size (100K--1M LOC); (2) high development activity (dense commit history and frequent CI runs); and (3) mature and stable CI/CD~\cite{cicdwiki} pipelines. These criteria are motivated by three considerations: large codebases with complex dependencies and build systems are more likely to surface repository-level failures and provide rich context for analyzing fixes; active development ensures sufficient volume and diversity of failures; and stable CI/CD pipelines reduce transient infrastructure noise and improve reproducibility. The selected projects are detailed in Section~\ref{sec:statistic}.

\subsection{Compilation Error Collection Pipeline}
Our goal in this stage is to recover \(\mathcal{E}\), a structured set of compilation diagnostics, from raw CI logs. Two difficulties arise in practice: CI logs are noisy and heterogeneous, and compilation errors can span multiple lines (\eg include stacks and notes). We address these challenges with two modules: (i) identify compilation-related failed CI jobs and retrieve their raw logs, and (ii) extract error candidates and enrich them with context to form structured error records. We then identify and minimize the corresponding developer repair patches in Section~\ref{sec:repair_patch} to construct complete error--fix instances.

\subsubsection{CI Failure Data Collection}\label{sec:ci_data_collect}
This module collects raw failure logs from GitHub Actions~\cite{actions}, which serve as the primary evidence for compiler diagnostics. We adopt a top-down refinement procedure to ensure that the failures are compilation-related and that the retrieved logs are attributable to a specific job:

\textit{Identifying Compilation-Related Workflows.}
We parse CI workflow YAML files to identify workflows that contain explicit compilation commands or build tool invocations (\eg \texttt{make}, \texttt{cmake}, \texttt{g++}, \texttt{clang++}, \texttt{ninja}~\cite{ninjawiki}).

\textit{Locating Failed Workflow Runs.}
Within a predefined time window (\eg the last 90 days), we retrieve runs of these workflows that are marked as \texttt{failed}.

\textit{Pinpointing the Failed Job.}
For each failed workflow run, we analyze its constituent jobs. By examining the execution status of each job, we pinpoint the specific job responsible for the failure.

\textit{Extracting Raw Logs.}
For each failed job, we download the complete raw log, and store it together with metadata (repository, workflow/job name, triggering commit SHA, branch, timestamp).

This structured storage decouples downstream analysis from the CI provider API and supports reproducible re-processing.

\subsubsection{Compilation Error Extraction}\label{sec:com_err_extra}
This module transforms raw job logs into structured compilation error records. The design goal is to be robust to log noise while maintaining high precision on true compiler diagnostics. Accordingly, we adopt a staged extraction strategy that first localizes candidate error lines, then reconstructs multi-line diagnostic context, and finally assigns an error taxonomy label.

\textit{Log Preprocessing and Normalization.} To handle potential encoding inconsistencies across different platforms, the pipeline first uses the \texttt{chardet} library~\cite{chardet} to detect each log file's encoding and normalizes it to UTF-8. Concurrently, common noisy content, such as CI system timestamps, is removed using regular expressions to streamline subsequent pattern matching.

\textit{Identifying Candidate Error Lines (Dual Filtering).}
A line is considered a candidate if it (i) does \emph{not} match any item in \texttt{EXCLUDE\_KEYWORDS} (\eg \texttt{test}, \texttt{warning:}, \texttt{clang-format}), and (ii) matches at least one item in \texttt{REQUIRED\_ERROR\_PATTERNS} (\eg \texttt{: error:}, \texttt{: fatal error:}, \texttt{undefined reference}). Candidates are further validated against \texttt{CPP\_ERROR\_PATTERNS}, which capture typical compiler formats (\eg \texttt{file.cpp:line:column: error:}).

\textit{Context Reconstruction.}
Since compiler diagnostics are often multi-line, we expand around each candidate to collect both (a) backward include traces (up to 20 lines, \eg \texttt{In file included from...}) and (b) forward diagnostic continuation (up to 15 lines, including \texttt{note:} lines and caret-marked snippets (\texttt{\^{}})). This process skips build-system noise (\eg \texttt{[  3\%] Building CXX object ...}) and terminates upon encountering the next distinct error or warning, ensuring the completeness and independence of each error's context.

\textit{Fine-Grained Error Taxonomy.}
To support subsequent analysis, we label each extracted error using a rule-based taxonomy. Our taxonomy contains \emph{eight} categories (Table~\ref{tab:error-categories}), including \textit{Warning as Error}. Because \textit{Warning as Error} is typically driven by compiler flags (\eg \texttt{-Werror}) and is often project-policy-specific, we exclude it from the released benchmark; therefore, the rest of the paper reports \emph{seven} categories.

Our \textit{Classification Method} matches each error message against a curated set of regular expressions distilled from a large corpus of real-world C/C++ compilation errors. To resolve ambiguities where one message matches multiple rules, we apply rules under a prioritized matching strategy, so that more specific and critical error types are identified first; the priority order is also given in Table~\ref{tab:error-categories}.

\begin{table}[]
\centering
\footnotesize
\caption{The Eight Major Error Categories in Our Taxonomy, Ordered by Matching Priority (\textit{Warning as Error} is excluded from the released benchmark).}
\label{tab:error-categories}
\begin{tabularx}{\linewidth}{@{}llX@{}}
\toprule
\textbf{Category} & & \textbf{Description} \\
\midrule
Warning as Error  & & Warnings that are elevated to errors by compiler flags (\eg \texttt{-Werror}). \\
Declaration Error & & Problems with undefined, redeclared, or conflicting identifiers (\eg symbols, namespaces). \\
Type Error        & & Violations of the C++ type system, such as mismatches, invalid conversions, or incomplete types. \\
Syntax Error      & & Issues related to code grammar and structure, like unexpected tokens or missing punctuation. \\
Template Error    & & Errors specific to C++ templates, such as instantiation or argument deduction issues. \\
Member Error      & & Issues with accessing or defining members of classes, including access specifier violations. \\
Function Error    & & Errors in function calls, overloads, constructors, or argument mismatches. \\
Semantic Error    & & Errors related to code logic and meaning, such as invalid operator usage or static assertions. \\
\bottomrule
\end{tabularx}
\vspace{-5pt}
\end{table}

\textit{Data Aggregation and Storage.}
To avoid double-counting repeated diagnostics triggered by multiple CI jobs for the same commit, we deduplicate errors at the commit level by merging jobs whose normalized diagnostic sets are identical. The diagnostic set is compared by keying on the extracted primary error messages and key fields such as file/line locations. All extracted information is stored in JSONL records with fields including \texttt{repo}, \texttt{commit\_sha}, \texttt{branch}, \texttt{timestamp}, \texttt{workflow\_name}, \texttt{job\_names}, \texttt{error\_lines}, \texttt{error\_details}, and \texttt{error\_types}.

\subsection{Repair Patch Identification}\label{sec:repair_patch}
Given a failure commit \(c_f\) and its diagnostic set \(\mathcal{E}\), the objective of this stage is to identify the corresponding developer repair commit \(c_r\) and to isolate a minimal patch \(\Delta_r\) that is plausibly responsible for eliminating the observed compilation failures. This is non-trivial because CI failures often occur on ephemeral feature branches, where the immediate subsequent commits may be unavailable after merges and branch deletion, and fixing commits frequently entangle unrelated changes (features or refactors) that would pollute ground truth. We address this with a multi-source candidate search, followed by validation and dependency-based minimization.

\subsubsection{Candidate Search and Ranking}
Starting from \(c_f\), we build a candidate set of potentially fixing commits using a \textit{multi-source search} strategy:
\begin{itemize}[leftmargin=*, labelsep=0.5em, itemsep=0.5ex, topsep=0.5ex, parsep=0ex]
    \item \textbf{Subsequent commits on the same branch:} commits after \(c_f\) on the originating branch (when the branch remains accessible), which typically corresponds to the most common repair path during local development.
    \item \textbf{Dependent commits across branches:} commits for which \(c_f\) is an ancestor, discovered via \texttt{git rev-list --ancestry-path}, covering merges, rebases, and cherry-picks. Since the candidate codebase directly evolves from the failure commit, this ancestor relation provides a causal relevance guarantee.
    \item \textbf{Same-author commits on the main branch:} commits by the same author on \texttt{main}/\texttt{master} after \(c_f\), covering PR-merged fixes.
\end{itemize}

After deduplication, the collected candidates are ranked using a multi-criteria heuristic algorithm. The ranking prioritizes candidates based on: (1) source type (\textit{same\_branch} > \textit{dependency} > \textit{main\_branch}), (2) temporal proximity to the failure (closer is better), and (3) change size (fewer changed files and lines are better). This strategy ensures that the most probable and relevant fixing commits are analyzed first.

\subsubsection{Validation and Filtering}
We apply three filters to isolate genuine fixes and extract a minimal patch.

\textit{CI Status Validation.}
We retain only candidates whose CI checks conclude success (\ie the corresponding workflow runs pass), to avoid choosing commits that merely shift the failure elsewhere.

\textit{Error File Deletion Check.}
We discard candidates that fix the failure by deleting the error-producing file, since such changes do not represent meaningful repair.

\textit{Dependency-Based Patch Minimization.}
For each remaining candidate, we compute the file-level diff and test whether the modified files intersect with the compilation dependency closure of the error-producing file. Let \(f\) be the primary error file extracted from \(\mathcal{E}\). We build a dependency closure \(D(f)\) by recursively parsing local \texttt{\#include} directives to collect directly/indirectly included headers reachable from \(f\). A candidate is considered a valid repair only if its modified file set \(M(c)\) satisfies \(M(c)\cap (\{f\}\cup D(f))\neq\emptyset\). We then define \(\Delta_r\) as the restriction of the commit diff to files in this intersection, ordered by proximity (\(f\) first, then direct includes, then transitive includes).

\subsubsection{Best Patch Selection}
We traverse validated candidates in ranking order and select \(c_r\) by a maximization principle: choose the commit that resolves the largest subset of the original errors in \(\mathcal{E}\). We use early termination: once a candidate eliminates all extracted errors, the search stops.

\subsubsection{Data Aggregation and Storage}
We store validated error--fix pairs as JSONL records with \texttt{failure\_commit}, \texttt{repair\_commit}, \texttt{error\_lines}, \texttt{error\_details}, \texttt{workflow\_name}, \texttt{job\_names}, and \texttt{compilation\_related\_files}. Within each record, diagnostics are deduplicated by error message to ensure that only unique errors are retained.

\subsection{Automated Reproduction Environment Construction}\label{sec:reproduce}
This stage constructs the reproduction artifact \(\mathcal{A}\) to replay the original failure and validate fixes. The core challenge is that a CI workflow YAML is a \emph{declarative} specification, while a failure occurs in a \emph{concrete} runtime environment with implicit parameters (runner image, matrix expansion, environment variables). We therefore reconstruct both the code state and the runtime state, and then infer the precise job configuration that produced the failure.

\subsubsection{State Restoration}
We restore the repository to the exact historical state by cloning and checking out \(c_f\) (\texttt{git checkout}).

\subsubsection{Environment Emulation}
We emulate the CI runtime with a two-layer Docker strategy. The \textit{base layer} maps logical runner labels (\eg \texttt{ubuntu-latest}) to version-locked public images (\eg \texttt{catthehacker/ubuntu:act-latest}), mitigating the inaccessibility of proprietary runner images. On top of this, the \textit{project-specific layer} executes a curated \texttt{PRE\_BUILD\_COMMAND} to install missing dependencies (\eg \texttt{apt-get install -y build-essential cmake}) and caches the result via \texttt{docker commit}. This turns a one-time manual reconstruction effort into a deterministic setup process for subsequent reproductions.

\subsubsection{Job Configuration Inference}
With code and environment in place, we infer the concrete job configuration and execute it via \texttt{act}~\cite{act}. The main difficulty is mapping the human-readable job name observed in CI logs (\eg \texttt{build (gcc, 13, mold)}) back to (i) the static job ID in YAML and (ii) the specific matrix assignment that instantiated that job. We perform three steps.

\textit{Intelligent Job ID Identification.}
We enumerate possible job instances by combining \texttt{act -l} with the workflow's matrix definitions. We then match the observed job name using a two-stage heuristic: exact prefix matching first, and a Levenshtein-distance-based similarity fallback when prefixes are ambiguous; we select the best-scoring instance under this similarity.

\textit{Matrix Parameter Inference.}
For \texttt{strategy: matrix} jobs, the execution is determined by a combination of variables. We parse the matrix keys and candidate values from YAML, tokenize the observed job name, and infer the concrete matrix assignment by maximizing token coverage. Specifically, for each matrix key, we choose the value whose string form appears in (or best matches) the tokenized job name; meanwhile, we validate the inferred key--value pairs against the enumerated instances from the previous step to avoid impossible combinations. The final assignments are materialized as \texttt{--matrix} flags for \texttt{act}.

\textit{Dynamic Command Synthesis and Execution.}
We synthesize a complete \texttt{act} command by assembling the workflow path (\texttt{-W}), the inferred job ID (\texttt{-j}), the trigger event (\eg \texttt{push}), the inferred \texttt{--matrix} parameters, runner-to-image bindings (\texttt{-P}), and required environment variables (\texttt{--env}, \eg \texttt{GITHUB\_TOKEN}). We redirect both stdout and stderr to log files for debugging and auditing.

\subsubsection{End-to-End Execution Verification}
Finally, we enforce an execution-level validity criterion for each candidate instance. We replay the \emph{same} failing job \(j_f\) at \(c_f\) under \(\mathcal{A}\) and require that it reproduces the extracted diagnostics. We then apply \(\Delta_r\) (a clean cherry-pick restricted to the selected files) and re-run \(j_f\) under the same environment. An instance is kept only if the job reaches a successful compilation outcome (\textit{fail--fix--pass}) and the originally extracted diagnostics no longer appear in the re-run logs. Only instances that satisfy this criterion are retained in the released \combench. All retained instances are further cross-validated by three experts to ensure quality.

%% file: sections/04-analysis.tex
\subsection{Data Statistics and Distribution}\label{sec:statistic}
This section presents a detailed data analysis of \combench, with the objective of demonstrating the representativeness of the benchmark of real-world, repository-level compilation errors. 
We first characterize the projects included in the benchmark (Table~\ref{tab:project-overview}), then provide an in-depth examination of the distribution of compilation error types (Figure~\ref{fig:error_distributions}). We also present the data collected in each phase to illustrate the progressive filtering and reduction of data (Table~\ref{tab:data_entry_phases}).

\subsubsection{Project Characteristics}
Our \combench benchmark collects errors from four large-scale C/C++ projects. Table~\ref{tab:project-overview} provides a detailed overview of these projects, including their main language, lines of code (LoC), number of code files, commits, workflow runs, and the total compilation errors identified. Specifically, for the LoC and file count metrics, only source files with common C/C++ extensions (\texttt{.c}, \texttt{.cpp}, \texttt{.cc}, \texttt{.h}, and \texttt{.hpp}) were considered.
As shown in Table~\ref{tab:project-overview}, the four projects collectively span over 1.4 million lines of code and 69,500 files. This substantial scale and complexity are representative of real-world software development. Among these, LLVM exhibits the highest number of commits and runs, consequently contributing the most errors (105) to the benchmark. 
A key observation from Table~\ref{tab:project-overview} is that each compilation error in our dataset has an average of 204 dependent files. This widespread dependency demonstrates that our benchmark's errors are not isolated, single-file issues but complex, repository-level problems. Consequently, \combench is well-suited to evaluate an ACER tool's ability to perform the global analysis required for real-world error resolution.
Furthermore, the benchmark includes a mix of C (OpenSSL) and C++ projects (Bitcoin, RocksDB, LLVM). This selection covers diverse application domains, such as security (OpenSSL), blockchain (Bitcoin), database management (RocksDB), and compiler infrastructure (LLVM). This diversity ensures that \combench accurately reflects a variety of real-world development
scenarios.
\begin{table}[]
\centering
\footnotesize 
\caption{Overview of Projects Analyzed in \combench.}
\label{tab:project-overview}
\begin{tabularx}{\linewidth}{@{}l l S[table-format=7,group-separator={,}] S[table-format=5,group-separator={,}] S[table-format=6,group-separator={,}] S[table-format=7,group-separator={,}] S[table-format=3,group-separator={,}] S[table-format=4.1]@{}}
\toprule
\textbf{Repository} & \textbf{Language} & {\textbf{LoC}} & {\textbf{Files}} & {\textbf{Commits}} & {\textbf{Workflow Runs}} & {\textbf{\makecell{Compilation \\ Errors}}} & {\textbf{\makecell{Avg. Dependent \\ Files per Error}}} \\
\midrule
Bitcoin~\cite{bitcoin}   & C++ & 376201   & 1452   & 46307   & 11388   & 12  & 47 \\
OpenSSL~\cite{openssl}   & C   & 217931   & 3809   & 37991   & 100788  & 19  & 13 \\
RocksDB~\cite{rocksdb}   & C++ & 724296   & 1314   & 13420   & 4706    & 64  & 24 \\
LLVM~\cite{llvm}      & C++ & 99585    & 62932  & 551028  & 1251657 & 105 & 366 \\
\midrule
\textbf{Total}   & & \textbf{1,418,014} & \textbf{69,507} & \textbf{648,746} & \textbf{1,368,539} & \textbf{200} & \textbf{204} \\
\bottomrule
\end{tabularx}
\end{table}

\subsubsection{Error Type Distribution}
Following the project characterization, we further analyzed the type distribution of all compilation errors collected across the four projects in our \combench benchmark. This characterization is crucial for two reasons: first, it demonstrates the breadth of error types covered by our benchmark, and second, it provides insights into the typical distribution of compilation errors encountered in real-world C/C++ development.

\begin{figure*}
    \centering
    \begin{subfigure}[b]{0.48\textwidth}
        \includegraphics[width=\linewidth]{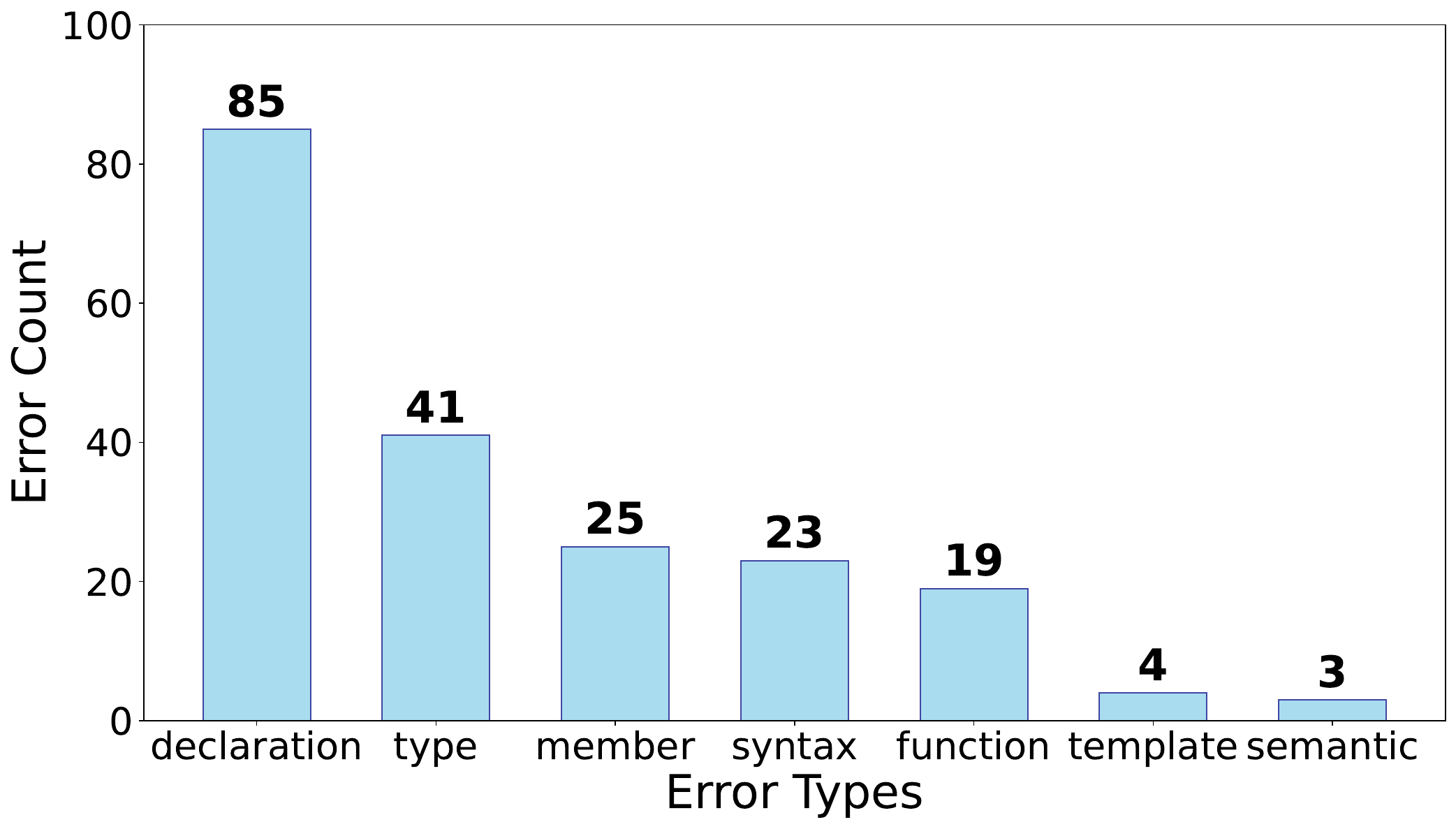}
        \caption{Error Distribution Across ALL Types}
        \label{fig:overall_error_distribution}
    \end{subfigure}
    \hfill
    \begin{subfigure}[b]{0.48\textwidth}
        \includegraphics[width=\linewidth]{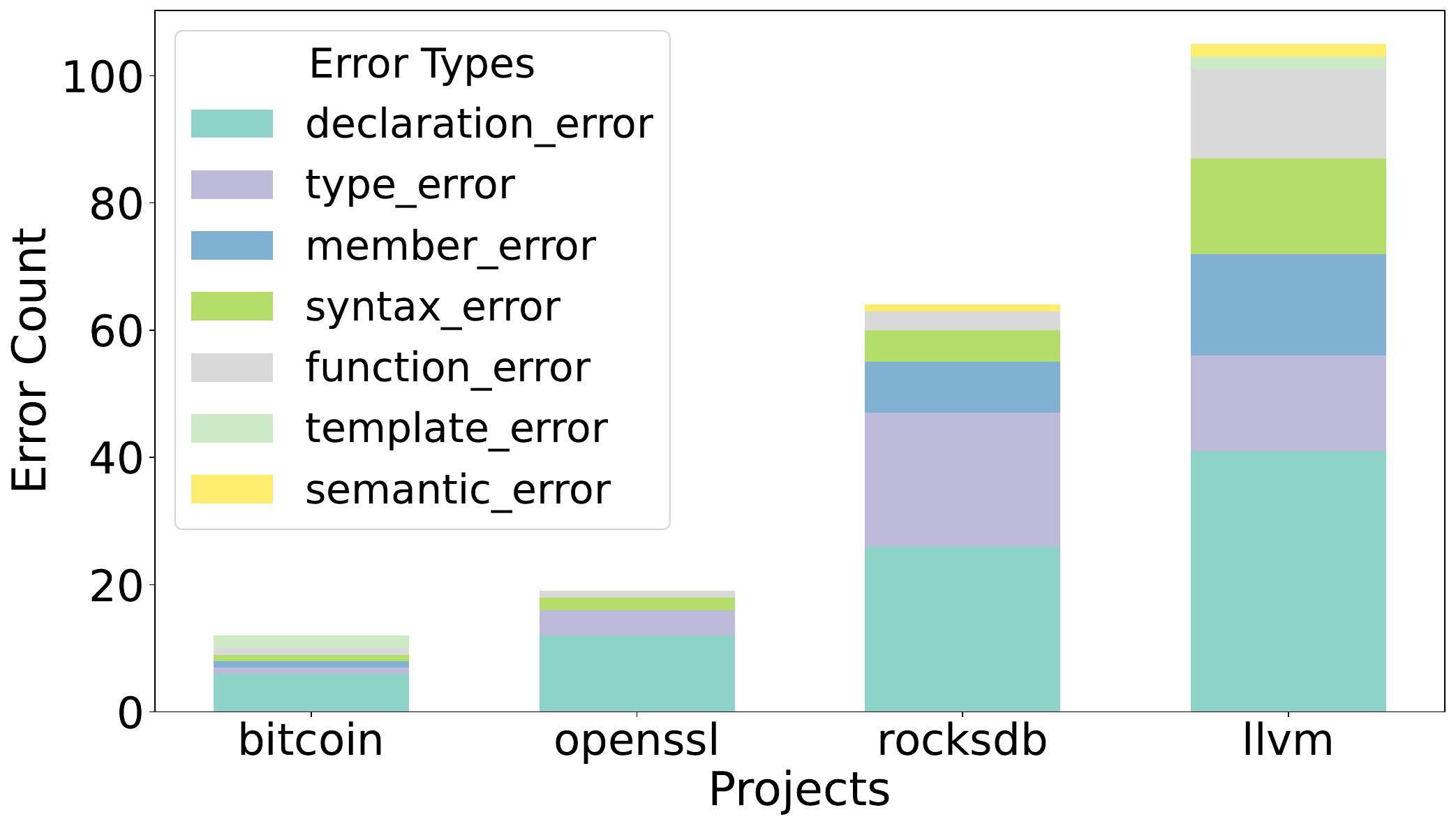}
        \caption{Error Type Distribution Across ALL Projects}
        \label{fig:project_error_comparison}
    \end{subfigure}
    \caption{Comparison of Error Distributions Across Projects and Error Types.}
    \label{fig:error_distributions}
    \vspace{-5pt}
\end{figure*}

Figure~\ref{fig:overall_error_distribution} illustrates the distribution of all 200 compilation errors across the identified categories. Our analysis reveals that declaration errors are the most prevalent, accounting for 85 instances. Following these, type errors (41), member errors (25), and syntax errors (23) represent the next most common categories. Less frequent, but still present, are function errors (19), template errors (4), and semantic errors (3). In total, we identified all seven main error categories within the benchmark, thereby demonstrating comprehensive coverage of common compilation issues.
We observe an error type skew, which is consistent with the natural frequency of compiler errors in real-world CI/CD pipelines. While this makes \combench more realistic, it may cause aggregate scores to be dominated by frequent categories. Accordingly, we report both overall and per-type results.

A cross-project analysis, detailed in Figure~\ref{fig:project_error_comparison}, further reveals a consistent pattern in the distribution of error types across projects. Across the four analyzed projects, the relative prevalence of error categories generally follows the overall trend: declaration errors are most frequent, followed by type errors, member errors, syntax errors, function errors, template errors, and semantic errors. This consistency suggests that these error types represent fundamental and pervasive compilation issues in C/C++ development, with their relative frequencies being largely independent of specific project characteristics.



\subsubsection{Data Entry Collected at Each Phase}
Table \ref{tab:data_entry_phases} provides an overview of the data collected at each distinct phase for every project within \combench. These phases track the data from the initial \textit{Raw Log for Failed Job}, through \textit{Extracted Compilation Error}, \textit{Repair Patch Identification}, and \textit{Deduplication}, culminating in \textit{Success Reproduction}.
As evident from the table, the number of entries generally decreases at each subsequent phase, reflecting the progressive filtering and refinement of the data. For instance, LLVM contributes the largest volume of initial raw logs (5,004) and extracted compilation errors (4,412), indicating its complexity and the richness of its error data. This high volume is then significantly reduced through subsequent stages, including deduplication, to yield 105 successful reproductions. In contrast, Bitcoin, starting with 585 raw logs, undergoes similar filtering, resulting in 42 deduplicated entries and ultimately 12 successful reproductions, suggesting a higher difficulty in achieving full reproduction for its issues. This detailed breakdown highlights the varying data availability and processing challenges across different projects.

\begin{table}[]
\footnotesize
\centering
\caption{Overview of Data Entry Collected at Each Phase for Each Project.}
\label{tab:data_entry_phases}
\begin{tabularx}{\linewidth}{@{}l >{\centering\arraybackslash}X >{\centering\arraybackslash}X >{\centering\arraybackslash}X >{\centering\arraybackslash}X@{} >{\centering\arraybackslash}X@{}}
\toprule
\textbf{Project} & \textbf{Raw Log for Failed Job} & \textbf{Extracted Compilation Error} & \textbf{Repair Patch Identification} & \textbf{Deduplication} & \textbf{Success Reproduction} \\
\midrule
Bitcoin~\cite{bitcoin} & 585 & 121 & 76 & 42 & 12 \\
OpenSSL~\cite{openssl} & 3938 & 162 & 27 & 19 & 19 \\
RocksDB~\cite{rocksdb} & 2240 & 265 & 146 & 86 & 64 \\
LLVM~\cite{llvm} & 5004 & 4412 & 168 & 115 & 105 \\
\midrule 
\textbf{Total} & \textbf{11,767} & \textbf{4,960} & \textbf{417} & \textbf{262} & \textbf{200} \\
\bottomrule
\end{tabularx}
\vspace{-10pt}
\end{table}

%% file: sections/05-evaluation.tex
\section{Experiment}
To comprehensively evaluate the ACER capabilities of LLMs, we designed an experimental framework. This framework employs three distinct metrics to assess the quality of generated patches, two experimental settings to probe different aspects of LLM performance (from pure patch generation to autonomous debugging), and 12 diverse models of varying sizes. This multi-faceted approach allows us to provide a holistic understanding of current LLM capabilities in ACER, highlighting their strengths and limitations across different repair scenarios and model scales.

\subsection{Evaluation Metrics}

We adopt three hierarchical metrics to evaluate generated patches, moving from basic syntax validity to deep semantic alignment.

\textbf{Compile Success (CS):} This metric measures whether the patch generated by the model successfully resolves the compilation error in the failing CI build without introducing new ones. However, some patches may achieve compilation success by inadvertently altering program semantics, which can lead to unexpected or incorrect program behaviors.

\textbf{Semantic Correctness (SC):} To address the risk of "shallow fixes" (\ie{} patches that compile but alter intended logic), SC evaluates whether the repair's semantics align with the developer's intent. It goes beyond mere compilation to assess the functional correctness of the fix. Notably, semantic inconsistency can arise for any type of compilation error, not only the \textit{semantic} error category. This metric is measured by manual analysis. 

\textbf{Exact Match (EM):} For patches that successfully compile, this metric examines whether the generated patch is subsumed by the ground-truth repair after normalization (\ie{} removing whitespace and comments). This stringent measure captures the model's ability to reproduce developer-level fixes. Specifically, an exact match is defined as a scenario where the ground-truth repair diff entirely subsumes the generated patch diff.

\subsection{Baselines}
While traditional ACER tools are valuable, we prioritize SOTA LLM baselines as they are the most directly comparable under our repository-level setting. We exclude traditional tools mainly because most prior methods are designed for file-level tasks; adapting them to repository-level, multi-file CI repair requires substantial engineering effort and they are expected to underperform. We therefore leave such comparisons out of scope in this paper.

\subsubsection{Setup 1: Repair with Perfect Fault Localization (PFL)}
This experimental setup is designed to isolate and evaluate the LLM's core patch generation capability. The input prompt is constructed to provide the model with complete local context, including: (1) the full source code of the failing file and modified files within ground-truth fix; (2) the full, raw compiler diagnostic messages containing error line with error message and error-producing location with file, line, and column numbers, alongside with error detail containing note and code reference (\texttt{\^{}}); and (3) a clear instruction to generate a patch. In this single-shot generation approach, the LLM is expected to output a single repair with cross-files multi-hunk fixes directly in a search-and-replace format, without multi-turn interaction. The evaluation, therefore, focuses purely on the model's effectiveness at generating a correct patch under these ideal input conditions. We retry one time if the LLMs provide no applicable search-replace patch.

\subsubsection{Setup 2: Agent-based Iterative Repair}
In contrast, this setup evaluates an LLM's capability to autonomously diagnose, localize, and repair compilation errors in an interactive, multi-turn environment. Specifically, we leverage the OpenHands~\cite{wang2025openhandsopenplatformai} framework to implement this agent-based setup. The LLM acts as an agent within this framework, beginning with a minimal prompt that provides only the raw compiler error log. From there, it must reason about the problem and explore the entire codebase to solve it. The agent interacts with the environment using a predefined set of tools that allow it to inspect files (\texttt{fetch}), execute shell commands for compilation or searching (\texttt{execute\_bash}), apply code modifications (\texttt{str\_replace\_editor}), manage its internal strategy (\texttt{think}, \texttt{task\_tracker}), and signal completion (\texttt{finish}). The agent iteratively uses these tools, observes the outcomes, and refines its approach until the error is resolved. This setup measures the LLM's comprehensive, end-to-end debugging skills, including its effectiveness in fault localization, strategic tool use, and final patch generation. We retry one time if the agent provides no legal diff.
Our configurations utilize version 0.55.0 of the OpenHands GitHub repository~\cite{openhands_git}, which was the latest version available at the time of writing this paper. Furthermore, the maximum number of iterations is set to 100, aligning with the default configuration of SWE-bench~\cite{jimenez2024swebench} in OpenHands.

\subsection{Models}
To comprehensively evaluate the capabilities of modern LLMs on the ACER task, we selected 12 popular models to serve as our baselines. Table~\ref{tab:models} presents their names, sizes, context lengths, and maximum output tokens.
For the closed-source models, we utilized their official APIs. Specifically, for GPT-5~\cite{openai_gpt5} and GPT-4.1~\cite{openai_gpt4_1}, we used the OpenAI APIs~\cite{openai_api}. For Claude 3.7 Sonnet and Claude 4 Sonnet~\cite{anthropic_claude4}, we used the Anthropic APIs~\cite{anthropic_api}. For Gemini 2.5 Pro~\cite{google_gemini}, we used the APIs provided by Google~\cite{google_gemini_api} to generate code solutions. For the open-source models, including DeepSeek-V3~\cite{deepseek_v3}, Qwen3-235B-A22b, Qwen3-30B-A3b, and the various Qwen3 models (4B, 8B, 14B, 32B)~\cite{qwen_series}, we leveraged the API provided by Deep Infra~\cite{deepinfra_api} to generate code solutions.
For all models, we use \texttt{temperature=0.2} to produce high-quality answers, and set \texttt{max\_tokens} to the models' maximum value to support long output, with prioritized addition of relevant files to fit in context for PFL.

\begin{table}[]
\centering
\caption{Baseline LLMs Used in Our Evaluation}
\label{tab:models}
\footnotesize
\setlength{\tabcolsep}{3.5pt}
\begin{tabular}{@{}l rcc @{\hspace{15pt}} l rcc@{}} 
\toprule
\multicolumn{4}{c}{\textbf{Open-source Models}} & \multicolumn{4}{c}{\textbf{Closed-source Models}} \\
\cmidrule(r){1-4} \cmidrule(l){5-8}
\textbf{Model} & \textbf{Size} & \textbf{Context} & \textbf{Output} 
& \textbf{Model} & \textbf{Size} & \textbf{Context} & \textbf{Output} \\
\midrule
DeepSeek-V3~\cite{deepseek_v3}          & 671B                  & 128K  & 8{,}192
  & GPT-5~\cite{openai_gpt5}              & -        & 272K      & 128{,}000      \\
Qwen3-235B-A22b~\cite{qwen_series}      & 235B (22B active)     & 128K  & 8{,}192
  & GPT-4.1~\cite{openai_gpt4_1}            & -        & 1M     & 32{,}768        \\
Qwen3-30B-A3b~\cite{qwen_series}        & 30B (3B active)       & 128K  & 8{,}192
  & Claude 4 Sonnet~\cite{anthropic_claude4}    & -        & 200K      & 16{,}384        \\
Qwen3 (4/8/14/32B)~\cite{qwen_series}   & 4/8/14/32B            & 128K  & 8{,}192
  & Claude 3.7 Sonnet~\cite{anthropic_claude37sonnet}  & -        & 128K      & 16{,}384        \\
                     &                       &      & 
  & Gemini 2.5 Pro~\cite{google_gemini}     & -        & 1M     & 64{,}000        \\
\bottomrule
\end{tabular}
\vspace{-10pt}
\end{table}

\subsection{Results} \label{sec:exp-results}
\subsubsection{Overall Performance}
The overall experiment results are shown in Table~\ref{tab:performance}, across all 12 models, 4 repositories, and two settings. Analyzed on them, we derive the following findings:

\textit{Finding 1: The ``Compile-but-Incorrect'' Gap is Systematic.}
Across all models, we observe a consistent drop-off from CS to SC, highlighting that satisfying compiler diagnostics is often easier than preserving repository-level semantics. This gap is most evident for top-tier models: GPT-5 achieves high CS rate (73\%), yet only a subset of these compilable patches are semantically correct (SC 41\%), and even fewer exactly reproduce the developer repair (EM 20\%). In our per-instance semantic comparisons of compilable patches, we group these CS-but-not-SC failures into five recurring types: violations of semantic contracts (API/return-value/nullability), configuration-semantic boundary violations (conditional compilation/feature semantics broken yet still compilable), deviations in protocol or error-handling paths that alter externally observable behavior, mismatched resource/policy/charging formulas that change runtime strategies, and alternative implementations that bypass repository abstractions and layering.

\textit{Finding 2: Performance Generally Scales with Model Size, but Plateaus in the Mid-range.}
There is a positive correlation between model size and repair capability, particularly within the Qwen3 family.
The largest model, Qwen3-235B, achieves a CS of 99.
This performance gradually decreases with model size, with the Qwen3-4B model achieving a CS of only 69.
However, performance appears to plateau among mid-sized models, with the 8B, 14B, 30B, and 32B variants exhibiting similar behavior across CS and SC. This suggests that, beyond a threshold, scaling alone is insufficient to close the CS--SC gap without better repository-level reasoning and convention awareness.

\textit{Finding 3: Models Exhibit Error-type Sensitivity.}
Beyond general performance rankings, we analyze each model's performance across our predefined error categories, presented in Figure~\ref{fig:types}. Different top-tier models show distinct strengths: Gemini-2.5-Pro is more effective on \textit{type} errors, while Claude-Sonnet-4 demonstrates stronger performance on \textit{member} errors. This indicates that ACER provides a fine-grained probe for different reasoning skills, thereby providing a clear roadmap for targeted future improvements.

\begin{table}[t]
\centering
\small 
\caption{Performance across different LLMs on \combench. We report Compile Success (CS), Semantic Correctness (SC), and Exact Match (EM). \textbf{Bold} indicates the best performance in each category.}
\label{tab:performance}
\setlength{\tabcolsep}{2.8pt} 
\begin{tabular}{@{}l ccc ccc ccc ccc >{\columncolor{graybg}}c>{\columncolor{graybg}}c>{\columncolor{graybg}}c @{}}
\toprule
\multirow{2}{*}{\textbf{Model}} 
& \multicolumn{3}{c}{\textbf{bitcoin (12)}} 
& \multicolumn{3}{c}{\textbf{openssl (19)}} 
& \multicolumn{3}{c}{\textbf{rocksdb (64)}} 
& \multicolumn{3}{c}{\textbf{llvm (105)}} 
& \multicolumn{3}{c}{\textbf{Total (200)}} \\
\cmidrule(lr){2-4} \cmidrule(lr){5-7} \cmidrule(lr){8-10} \cmidrule(lr){11-13} \cmidrule(l){14-16}
& CS & SC & EM & CS & SC & EM & CS & SC & EM & CS & SC & EM & \textbf{CS} & \textbf{SC} & \textbf{EM} \\
\midrule
\textit{Proprietary Models} \\
GPT-5              & \textbf{9} & \textbf{3} & 1 & \textbf{15} & 8 & 6 & \textbf{41} & \textbf{23} & 12 & \textbf{81} & \textbf{48} & 21 & \textbf{146} & \textbf{82} & 40 \\
GPT-4.1            & 4 & 2 & 0 & 12 & 7 & 6 & 37 & 20 & 12 & 61 & 35 & 15 & 114 & 64 & 33 \\
Claude-Sonnet-4    & 7 & \textbf{3} & 1 & 12 & 8 & 6 & 36 & 22 & \textbf{13} & 59 & 35 & 21 & 114 & 68 & 41 \\
Claude 3.7 Sonnet  & 7 & 1 & 0 & 10 & 5 & 5 & 35 & 19 & 11 & 55 & 32 & 16 & 107 & 57 & 32 \\
Gemini-2.5-Pro     & 3 & 2 & 0 & 14 & \textbf{11} & \textbf{11} & 34 & 16 & 11 & 65 & 44 & \textbf{26} & 116 & 73 & \textbf{48} \\
\addlinespace[0.5em]
\textit{Open-source Models} \\
DeepSeek-V3        & 5 & 2 & 1 & 12 & 6 & 4 & 28 & 13 & 10 & 61 & 34 & 14 & 106 & 55 & 29 \\
Qwen3-235B-A22b    & 5 & 1 & 1 & 13 & 8 & 6 & 23 & 10 & 7 & 58 & 30 & 12 & 99  & 49 & 26 \\
Qwen3-30B-A3b      & 4 & 2 & 1 & 9 & 6 & 5 & 20 & 9 & 5 & 48 & 23 & 15 & 81  & 40 & 26 \\
Qwen3-32B          & 2 & 2 & 1 & 12 & 4 & 4 & 19 & 9 & 6 & 47 & 27 & 11 & 80 & 42 & 22 \\
\midrule
\textit{Agentic Framework} \\
Claude-Sonnet-4 \dag & 6 & 3 & 1 & 13 & 8 & 8 & 40 & 21 & 14 & 72 & 52 & 29 & 131 & 84 & 52 \\
Qwen3-32B \dag       & 1 & 1 & 0 & 9 & 4 & 5 & 21 & 10 & 7 & 43 & 23 & 8 & 74 & 38 & 20 \\
\bottomrule
\multicolumn{16}{l}{\footnotesize \dag~Evaluated within the OpenHands agentic framework.} \\
\end{tabular}
\vspace{-10pt}
\end{table}

\subsubsection{Agent vs. PFL}
Overall, within the OpenHands agentic framework, Claude-Sonnet-4 achieves notably higher CS and SC than PFL (see Table~\ref{tab:performance}), indicating that iterative exploration and context gathering can yield better end-to-end repair outcomes for strong models. At the type level, the agent improves performance on \textit{declaration}, \textit{type}, \textit{syntax}, and \textit{function} errors (especially in CS/SC), but drops on \textit{member} errors; for low-count categories such as \textit{template} and \textit{semantic}, the difference is not pronounced. Meanwhile, the agentic paradigm imposes a capability threshold: for Qwen3-32B, overall performance under the agentic framework decreases compared to PFL, suggesting that smaller models are more susceptible to noise and trial-and-error costs in retrieval, planning, and tool use. Our analysis, incorporating specific case studies and statistical data, is summarized as follows.

\begin{figure*}
    \centering
    \includegraphics[width=0.96\textwidth]{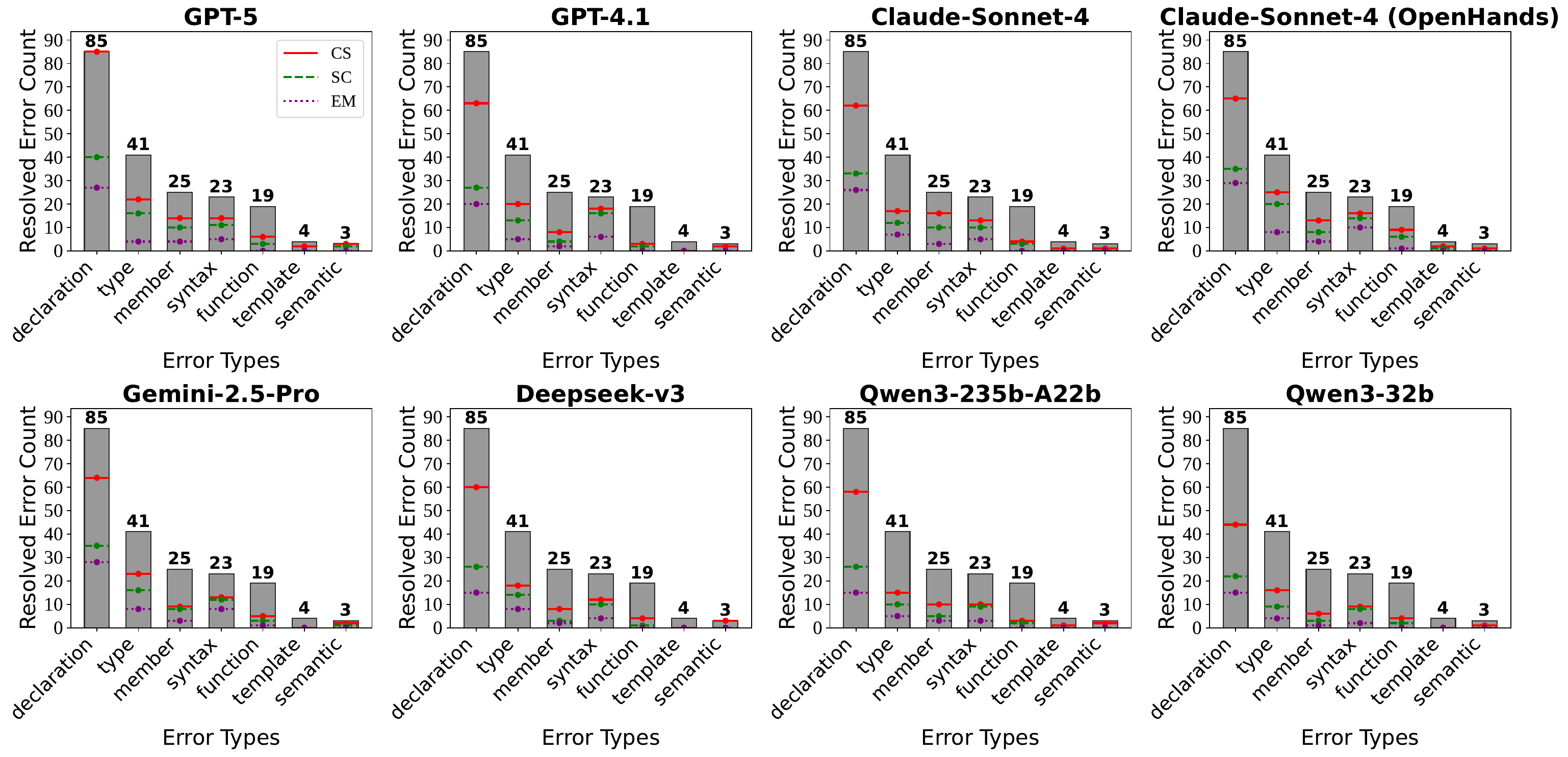}
    \caption{Performance of All LLMs Across Different Types}
    \Description{The bar chart compares the number of compilation errors resolved by each LLM across various types, highlighting differences in performance.}
    \label{fig:types}
    \vspace{-10pt}
\end{figure*}

\textit{Finding 4: Agentic Repair Improves Overall Performance, but the Gains Are Capability-dependent.}
Our analysis suggests that PFL and the agentic paradigm represent two complementary repair mechanisms. With perfect fault localization, PFL favors fast, low-overhead local edits around a single reported error site; the agent, through iterative retrieval and multi-turn validation, can integrate scattered repository-level context and thus better handle global repairs that require cross-file dependency reasoning and multi-location consistency. For example, when fixing a malformed ternary operator \texttt{(expression1 ? expression2)}, PFL may simplify the code into a logical \texttt{\&\&} expression, while the agent can preserve the original structure by adding the missing \texttt{: false} branch, reflecting different repair styles that prioritize conciseness versus completeness. For Claude-Sonnet-4, this extra context and validation yields higher CS/SC/EM (see Table~\ref{tab:performance}) and resolves more errors at the type level (see Figure~\ref{fig:types}). However, the gains are not uniform: some categories can still drop, indicating that iterative exploration can introduce mis-retrieval and trial-and-error costs; moreover, agentic repair demands stronger planning, retrieval, and tool-use competence, and may degrade for weaker models (\eg Qwen3-32B). Overall, the agentic framework can deliver better end-to-end repair performance for strong models, but its gains depend on the model's agentic capability and execution stability.

\textit{Finding 5: Some Hard Compilation Errors Remain Unsolved Even for Top-tier Models.}
In the PFL setting, the CS-failure intersection of GPT-5, Gemini-2.5-Pro, and Claude-Sonnet-4 still contains 34 compilation errors. These hard cases are rarely single-line fixes; instead, they often require cross-file and cross-configuration consistency. We summarize them into four recurring patterns: (1) API surface evolution that causes declaration--definition--call-site inconsistencies; (2) access-control and encapsulation constraints (\eg protected/private members or constructors) that restrict feasible fixes; (3) the complexity of template instantiation and overload resolution / include dependencies that complicates type visibility and alignment; and (4) coupled, cascading errors that require coordinated fixes along call chains or across translation units.

\subsection{Case Study}
We conducted a detailed case study on instances where the best-performing model, GPT-5, in the PFL setting generated patches that compiled successfully but were semantically incorrect. We group these failures into two recurring patterns.

\textbf{Failure Mode 1 (Violation of Repository Conventions).}
\textit{Case Study 1 (Bitcoin): Custom vs. Standard Library Mismatch.}
A common LLM failure mode is proposing a locally plausible fix that violates repository conventions. For an error, \texttt{'std::span' has not been declared}, the LLM's patch added the standard header \texttt{\#include <span>}. In contrast, the developer's ground-truth patch replaced a forward declaration (\texttt{- class Span;}) with a project-specific header (\texttt{+ \#include <span.h>}).
The LLM's patch, while syntactically plausible, is semantically incorrect. It introduces the standard C++20 \texttt{<span>} into a project designed around a custom, C++17-compatible \texttt{<span.h>}, violating its architectural constraints. This choice is incompatible with the project's custom data structures (\eg{} \texttt{prevector}) and can trigger compilation failures in other translation units or build configurations beyond the failing CI target. This case highlights that without a global understanding of a project's internal APIs, an LLM may default to a common but contextually invalid solution.

\textbf{Failure Mode 2 (Shallow Substitution).}
\textit{Case Study 2 (Openssl): Global Semantic Refactoring vs. Substitute Undefined Identifier with Arbitrarily Defined One.}
This case illustrates a common but flawed LLM repair strategy: resolving an undeclared identifier error by substituting the unknown identifier with an arbitrary, but defined, one from the nearby codes, without regard for semantic correctness. Faced with the undeclared error constant \texttt{PROV\_R\_FAILED\_DURING\_DERIVATION}, even with the clear hint in comments \texttt{/* the public key was probably a weak key */}, the LLM's patch still replaced it with a nearby, generic constant, \texttt{ERR\_R\_INTERNAL\_ERROR}, simply to satisfy the compiler. The developer's ground-truth patch, however, performed a deeper semantic correction, refactoring the entire error-handling block to use a more protocol-appropriate alert (\texttt{+ SSLfatal(..., SSL\_R\_BAD\_KEY\_SHARE)}). The LLM's substitution, while compilable, is semantically incorrect as it alters the program's externally observable behavior by sending a different and misleading TLS alert. This highlights the model's tendency to prioritize syntactic validity---finding any defined identifier of the correct type---over understanding the code's domain-specific logic.

%% file: sections/06-conclusion.tex



\section{Threats to Validity}

\textbf{Internal Threats.}
One internal threat lies in the dataset construction process.
There is a possibility that extracted patches contain unrelated code or that certain compilation errors are misclassified during CI log extraction.
To mitigate this risk, we apply multi-step validation and filtering processes to reduce noise.
%
Another internal threat is the reliance on compile success as a primary correctness signal, which does not necessarily guarantee semantic equivalence.
To address this, we introduced additional evaluation metrics, namely semantic correctness and exact match, and multiple experts were involved in the manual inspection of cases to ensure reliable assessments.

\textbf{External Threats.}
While \combench covers C/C++ projects, it cannot fully represent other programming languages.
Nevertheless, our design can be readily extended to other languages (\eg{} Java or Rust), supporting broader generalizability in future studies.
Another external threat is the non-deterministic nature of LLMs: the same input may yield different outputs.
We mitigate this by providing detailed inference configurations for LLMs, which helps reduce variability and improve reproducibility.

\section{Related Work}

\subsection{Automated Dataset Construction and Mining Frameworks}
Beyond ACER-specific benchmarks, software engineering has established various automated dataset construction and mining paradigms~\cite{bugminer, szz, tangLLM4SZZEnhancingSZZ2025, jimenez2024swebench, jiang2022bugbuilder, song2022regminer, just2014defects4j}. While these frameworks provide systematic ways to collect bug-fix pairs, their design choices differ from \combench in ways that matter for compilation failures in modern CI/CD.

First, regarding data visibility, most prior pipelines mine permanent artifacts such as commit histories or PR links~\cite{bugminer,szz,jiang2022bugbuilder,song2022regminer,jimenez2024swebench}. This introduces survivorship bias: it mainly captures bugs that survived the development process and were eventually merged. Consequently, it misses pre-merge compilation failures blocked by CI gatekeeping. \combench addresses this by mining transient CI/CD logs to capture such failures that never enter the permanent repository history.

Second, in terms of the verification oracle, existing benchmarks typically rely on test-driven oracles (e.g., SWE-bench and Defects4J~\cite{jimenez2024swebench,just2014defects4j}) and assume that a functional execution environment is already provided. In contrast, CI compilation failures are often environment-sensitive and require specific toolchains, dependencies, and build flags, which typically must be inferred and reconstructed from CI logs and configuration. \combench adopts a construction-as-validation principle: it treats the executable fail-fix-pass compile cycle as the oracle and validates repairs in a reconstructed Docker environment, ensuring that the repair is executable under that environment. 

\subsection{Existing ACER Benchmarks}
As discussed in Section~\ref{sec:benchmarks}, widely used ACER benchmarks such as DeepFix~\cite{10.5555/3298239.3298436} and TRACER~\cite{8445185} are constructed from decontextualized, single-file programs, typically from student assignments. This setting encourages local, snippet-level repairs with limited repository context, and it does not reflect CI failures that involve cross-file dependencies. In contrast, \combench provides repository-level snapshots, real CI-triggered compilation failures, and developer-provided fixes, enabling evaluation in complex, real-world scenarios.

\section{Conclusion}
In this paper, we construct \combench, a repository-level, real-world benchmark with comprehensive error types, reproducible environment, and a reusable data collection pipeline, which enables continuous evolving. Collected from 4 large-scale C/C++ projects across different domains, covering all 7 main categories of compilation errors, our benchmark is able to represent the real-world development scenario. Our benchmark evaluates both the compilable and sematic correct of the ACER tool's generated patches. These three metrics we designed reveal the error repair capabilities of each model from distinct perspectives and provide valuable insights and a solid evaluation foundation for future works in the ACER area.
